\documentclass[conference]{IEEEtran}
\IEEEoverridecommandlockouts
\usepackage{graphicx}
\usepackage{cite}
\usepackage{graphicx}
\usepackage[utf8]{inputenc}
\usepackage{color}
\usepackage{bm}
\graphicspath{{Figures/}}

\usepackage{booktabs}
\usepackage{amsmath}
\usepackage{amsthm}
\newtheorem*{remark}{Remark}
\usepackage{siunitx}
\DeclareSIUnit \pu {pu}

\usepackage{xcolor}

\usepackage{soul}

\definecolor{redFG}{rgb}{0.875,0,0.5}
\definecolor{blueFG}{rgb}{0,0.5,0.875}

\newcommand{\re}{{\rm Re}}
\newcommand{\im}{{\rm Im}}
\newcommand{\vm}{\mathrm{v}}
\newcommand{\sg}{\mathrm{s}}
\newcommand{\bus}{\mathrm{b}}
\newcommand{\load}{\mathrm{L}}
\newcommand{\ov}{\mathrm{v}}
\newcommand{\os}{\mathrm{s}}
\newcommand{\pv}{\mathrm{p,v}}
\newcommand{\qv}{\mathrm{q,v}}

\newcommand{\ptoomega}{$P\to\omega$}
\newcommand{\ptov}{$P\to V$}
\newcommand{\qtoomega}{$Q\to\omega$}
\newcommand{\qtov}{$Q\to V$}

\newcommand{\mt}{\mathrm}

\usepackage{hyperref}
\allowdisplaybreaks

\def\PlotScalingFactor{1.015}

\begin{document}

\title{The role of VSG parameters\\ in shaping small-signal SG dynamics\looseness=-1
}

\author{\IEEEauthorblockN{Ferdinand Geuss}
	\IEEEauthorblockA{\textit{University of Stuttgart} \\
		Stuttgart, Germany\\
		ferdinand.geuss@isys.uni-stuttgart.de}
		\and
		\IEEEauthorblockN{Orcun Karaca}
		\IEEEauthorblockA{\textit{ABB Corporate Research} \\
			Baden-Dättwil, Switzerland \\
			orcun.karaca@ch.abb.com}
	\and
	\IEEEauthorblockN{Mario Schweizer}
	\IEEEauthorblockA{\textit{ABB Corporate Research} \\
		Baden-Dättwil, Switzerland \\
		mario.schweizer@ch.abb.com} 
		\and 
	\IEEEauthorblockN{Ognjen Stanojev}
	\IEEEauthorblockA{\textit{ABB Corporate Research} \\
		Baden-Dättwil, Switzerland \\
		ognjen.stanojev@ch.abb.com}
}

\maketitle

\begin{abstract}
We derive a small-signal transfer function for a system comprising a virtual synchronous generator (VSG), a synchronous generator (SG), and a load, capturing voltage and frequency dynamics. Using this model, we analyze the sensitivity of SG dynamics to VSG parameters, highlighting trade-offs in choosing virtual inertia and governor lag, the limited effect of damper-winding emulation, and several others.  \looseness=-1
\end{abstract}


\section{Introduction}

As power electronic interfaces become more prevalent, particularly with the renewable integration, the industry increasingly requires advanced grid-forming (GFM) control. These controllers can form voltage, emulate inertia, and realize droop behavior. 
For such controllers, small-signal methods are widely used especially in single grid-connected converter systems,~e.g., with passivity or Nyquist criteria~\cite{harnefors2015passivity}. 
In contrast, this paper studies the small-signal properties of an interconnected system comprising a converter operated as a virtual synchronous generator (VSG)\footnote{Referring primarily to droop control, inertia, and damper winding emulation functionalities present in GFM.}, a synchronous generator (SG), and a load. Unlike the related studies in~\cite{Tayyebi2020,sun2021stability,Liu2022,pogaku2007modeling}, the current control and the network dynamics are neglected to focus only on slower dynamics. Such simplified models can reveal behaviors that more detailed models may obscure. This same motivation led to the foundational work in~\cite{liu2016vsg_small_signal}, which deliberately overlooked network dynamics to derive a parallel VSG–SG model, unveiling new facets of VSG parametrization. However, that study omits voltage and reactive power dynamics, does not analyze zero locations or their impact, and primarily highlights the benefits of higher inertia/damping and reduced governor delay. These limitations motivate our work, and we address each explicitly.

Among more detailed models,~\cite{Tayyebi2020} considers simulation studies for a nine-bus system, including the dc-link and current control dynamics, highlighting the interaction between fast GFM control and slow SG dynamics.
A small-signal study of parallel VSGs in~\cite{sun2021stability} determines their damping coefficients. 
In~\cite{Liu2022}, a small-signal model of a VSG-SG interconnection with network dynamics is derived to study low-frequency oscillations using pole plots. They state that increasing the governor lag and inductance could weaken the influence of the VSG.  

Related simplified models are found in~\cite{Lin2017,lasseter2019grid,Wu2016,DArco2016,li2016self}.
For example,~\cite{Lin2017} analyzes instability with grid-following (GFL) penetration; \cite{lasseter2019grid} shows damping versus penetration levels; \cite{Wu2016} derives a VSG small-signal model highlighting the coupling of QV and P$\omega$ dynamics; \cite{DArco2016} studies a GFL converter connected to a VSG; and~\cite{li2016self} examines inertia and damping effects in different time intervals during a frequency transient.


The main contribution of this paper is the derivation of a small-signal transfer function mapping load power to internal frequencies and voltages, which we use to analyze the interconnected system with respect to VSG parametrization, in particular, to study the sensitivity of SG dynamics. Compared with~\cite{liu2016vsg_small_signal}, the transfer function also captures voltage and reactive power dynamics, and we give specific attention to the locations of its zeros. Section~\ref{ch:Sensitivity} shows that these zero locations can significantly affect system behavior. In our analysis, we vary VSG parameters---inertia, damper winding constant, governor lag time constant, QV-droop gain and filter time constant, stator inductance, and XR-ratio---one at a time while keeping others at nominal values, to provide insights. These insights either confirm or extend prior findings. For example, we confirm the results of~\cite{Liu2022} (discussed above), further distinguish oscillations on two distinct time scales, and examine the influence of zeros in both active and reactive power dynamics. Compared with~\cite{DArco2016}, we also confirm the benefits of increased inertia, but we further highlight the advantages of inertia matching. Finally, we validate the transfer function and the main findings through simulation case studies.

\section{Preliminaries}
\label{ch:Preliminaries}

Parameters are denoted with upper-case, e.g., $K$. Bold capitals, e.g.,~$\bm{K}$, refer to parameter matrices, and~$K_{ij}$ describes the entry in the $i$th row and $j$th column of~$\bm{K}$. The identity is $\bm{I}$.
Functions in upper-case refer to transfer functions, e.g.,~$G(s)$, while transfer function matrices are bold, e.g.,~$\bm{G}(s)$. Here, $s$ is to be interpreted as the derivative operator.
Scalar variables are expressed by lower-case, such as~$v$, and their vector versions are denoted bold, e.g.,~$\bm{v}$.
Phasors are marked by an arrow, e.g., $\vec{v} = v\cdot e^{j\varphi} = v\angle \varphi$.
Subscripts $_\vm$, $_\sg$, $_\bus$ are used to distinguish quantities belonging to the VSG, the SG, and the bus, respectively.
Similarly, references are indicated by the subscript~$_\mt{r}$. 
For example, the reference frequency is $\omega_\mt{r}$.
For brevity, we often define quantities only for the VSG or only for the SG. Next, we describe the dynamics and the assumptions, and present the interconnected system. \looseness=-1

%
%

\subsection{VSG Modeling}
\begin{figure}[t]
	\centering
	\resizebox{.30\textwidth}{!}{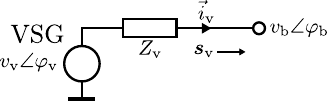}
	\caption{Diagram of the VSG model.}
	\label{fig:VSG_diagram}
\end{figure}

The VSG dynamics follow~\cite{liu2016vsg_small_signal}, with the difference that we include voltage dynamics (see Fig.~\ref{fig:VSG_diagram}). As in~\cite{liu2016vsg_small_signal}, the SG has the same structure as the VSG, with different parametrization. The VSG is modeled as a three-phase voltage source with magnitude $v_\vm$ and instantaneous phase angle $\varphi_\vm$ (i.e., with frequency $\omega_\vm = s\varphi_\vm$) in series with a virtual stator impedance~$Z_\vm$. 
In contrast to the pure inductances in \cite{liu2016vsg_small_signal}, we have resistive-inductive impedances, i.e., $Z_\vm = R_\vm + j\omega_\mathrm{r} L_\vm = R_\vm + jX_\vm$.
Small-signal changes in the reactance are similarly ignored.
The VSG is connected to a bus with magnitude $v_\bus$ and instantaneous phase angle $\varphi_\bus$ (i.e., $\omega_\bus=s\varphi_\bus$), and the complex power at the line connecting to this bus is denoted by $\bm{s}_\ov=[p_\ov\, q_\ov]^\top$ in vector-form, where $p_\ov$ and $q_\ov$ refer to active and reactive power, respectively. \looseness=-1

\subsubsection{Frequency Dynamics}
The dynamics of $\omega_\vm$ are governed by the swing equation, i.e.,
\begin{equation}
    M_\vm s \omega_\vm = \frac{p_\mt{v,r} - K_\pv\left(\omega_\vm-\omega_\mt{r}\right)}{1+T_\pv s}  - p_\ov - D_\vm\left(\omega_\vm-\omega_\bus\right)
    ,
    \label{eq:swing_eq_LS}
\end{equation}
with angular momentum $M_\vm$, assuming $M_\vm = J_\vm \omega_\mt{r}$. 
$J_\vm = \frac{2H_\vm}{\omega_\mt{r}^2}S_\mt{n}$ refers to the moment of inertia,
where $H_\vm$ and $S_\mt{n}$ are the inertia time constant and the nominal power, respectively.

The first term on the right-hand side of \eqref{eq:swing_eq_LS}, including the power setpoint $p_\mt{v,r}$ and the frequency droop with constant $K_\pv$, describes the speed governor. To model the typical delay of the prime mover, a low-pass filter with time constant $T_\pv$ is included. For a VSG, $T_\pv$ can be lower than that of a real SG. The third term approximates the effect of the damper windings. It is dependent on the difference between internal frequency and bus frequency through a damper winding constant~$D_\vm$.\looseness=-1


\subsubsection{Voltage Dynamics}
The voltage dynamics are given by
\begin{equation}
    v_\vm = v_\mt{v,r} + \left({K_\qv}/({1+T_\qv s})\right) \left(q_\mt{v,r} - q_\ov\right)
    .
    \label{eq:voltage_dynamics_LS}
\end{equation}
The internal voltage magnitude $v_\vm$ is determined by QV-droop, i.e., linear around its setpoint $v_\mt{v,r}$ based on the difference between $q_\ov$ and its setpoint $q_\mt{v,r}$ via the droop constant $K_\qv$.
Similar to the speed governor, the QV-droop is also subject to delays, modeled by a low-pass filter with time constant $T_\qv$.


\subsubsection{Power Flow}
The quasi-steady state output power supplied to the bus node with voltage $\vec{v}_\bus$ is given by 
{\medmuskip=.05mu\thickmuskip=.05mu\thinmuskip=.05mu
\begin{align}
    p_\ov &= \re\{\vec{v}_\bus \vec{i}_\vm^*\} = \frac{R_\vm\left(v_\vm v_\bus \cos(\theta_\vm) - v_\bus^2\right) + X_\vm v_\vm v_\bus \sin(\theta_\vm)}{R_\vm^2+X_\vm^2}
    ,
    \label{eq:power_flow_active_LS}
    \\
    q_\ov &= \im\{\vec{v}_\bus \vec{i}_\vm^*\} = \frac{X_\vm \left(v_\vm v_\bus \cos(\theta_\vm) - v_\bus^2\right)-R_\vm v_\vm v_\bus \sin(\theta_\vm)}{R_\vm^2+X_\vm^2},
    \label{eq:power_flow_reactive_LS}
\end{align}}where $\theta_\vm = \varphi_\vm - \varphi_\bus$ denotes the angle difference over the stator impedance, also satisfying $s\theta_\vm = \omega_\vm - \omega_\bus$.

%
%

\subsection{Setup of the Interconnected System}
\label{ch:system_setup}

\begin{figure}[t]
    \centering
    \resizebox{0.41\textwidth}{!}{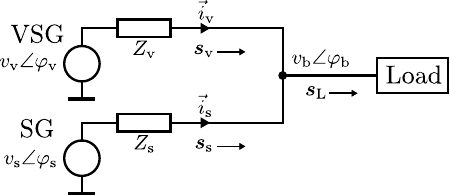}
    \caption{Diagram of the  interconnected system model.}
    \label{fig:system_diagram}
\end{figure}

The VSG, SG, and load are directly connected to a common bus (Fig.~\ref{fig:system_diagram}).The SG and VSG share the same dynamics with different parametrization, and the load power is
$\bm{s}_\load = \bm{s}_\ov + \bm{s}_\os$. Line impedances between the VSG and SG are assumed negligible. Modeling them would require distinct bus nodes and additional equations, which we relegate to future work.

\section{Small-Signal Dynamics}
\label{ch:Dynamics}

We first derive a small-signal version of the VSG model, followed by its stand-alone operation and the complete system.
For any $x$, let $x = x_0 + \Delta x$, where $x_0$ represents the steady-state value (i.e., operating point), while $\Delta x$ denotes the perturbation variable.
Small-signal representation of \eqref{eq:swing_eq_LS} is:
{\medmuskip=.05mu\thickmuskip=.05mu\thinmuskip=.05mu\begin{equation}
    M_\vm s \Delta\omega_\vm = -\left({K_\pv}/({1+T_\pv s})\right)\Delta\omega_\vm - \Delta p_\ov - D_\vm\left(\Delta\omega_\vm-\Delta\omega_\bus\right).
    \label{eq:swing_eq_ss}
\end{equation}}Similarly, the small-signal representation of \eqref{eq:voltage_dynamics_LS} is 
\begin{equation}
    \Delta v_\vm = -\left({K_\qv}/({1+T_\qv s})\right) \Delta q_\ov
    .
    \label{eq:voltage_dynamics_ss}
\end{equation}

Notice that \eqref{eq:power_flow_active_LS} and \eqref{eq:power_flow_reactive_LS} are nonlinear. To this end, we linearize \eqref{eq:power_flow_active_LS} and \eqref{eq:power_flow_reactive_LS} with respect to $\theta_\vm$, $v_\vm$, and~$v_\bus$.
Separating operating points and perturbation variables, we get
%
\begin{equation}
    \Delta\bm{s}_\ov =
    [\Delta p_\ov \; \Delta q_\ov]^\top
    = \bm{K}_\vm 
    [\Delta\theta_\vm \; \Delta v_\vm \; \Delta v_\bus]^\top
    ,
    \label{eq:power_flow_ss}
\end{equation}
with the power flow matrix $\bm{K}_\vm$ as in \eqref{eq:linearized_power_flow_matrix} (provided on top of the next page).
\begin{figure*}

{\medmuskip=.12mu\thickmuskip=.12mu\thinmuskip=.12mu
\begin{equation}
    \bm{K}_\vm = 
    \begin{bmatrix}
        \frac{-R_\vm v_\mt{v,0} v_\mt{b,0} \sin(\theta_\mt{v,0}) + X_\vm v_\mt{v,0} v_\mt{b,0} \cos(\theta_\mt{v,0})}{R_\vm^2+X_\vm^2}
        &
        \frac{R_\vm v_\mt{b,0} \cos(\theta_\mt{v,0}) + X_\vm v_\mt{b,0} \sin(\theta_\mt{v,0})}{R_\vm^2+X_\vm^2}
        &
        \frac{R_\vm \left(v_\mt{v,0} \cos(\theta_\mt{v,0}) - 2v_\mt{b,0}\right) + X_\vm v_\mt{v,0} \sin(\theta_\mt{v,0})}{R_\vm^2+X_\vm^2}
        \\
        \frac{-R_\vm v_\mt{v,0} v_\mt{b,0} \cos(\theta_\mt{v,0}) - X_\vm v_\mt{v,0} v_\mt{b,0} \sin(\theta_\mt{v,0})}{R_\vm^2+X_\vm^2}
        &
        \frac{-R_\vm v_\mt{b,0} \sin(\theta_\mt{v,0}) + X_\vm v_\mt{b,0} \cos(\theta_\mt{v,0})}{R_\vm^2+X_\vm^2}
        &
        \frac{-R_\vm v_\mt{v,0} \sin(\theta_\mt{v,0}) + X_\vm \left(v_\mt{v,0} \cos(\theta_\mt{v,0}) - 2 v_\mt{b,0}\right)}{R_\vm^2+X_\vm^2}
    \end{bmatrix}
    \label{eq:linearized_power_flow_matrix}\vspace{.1cm}
\end{equation}
}%
\hrule
\vspace{.1cm}
\end{figure*}
Finally, we have 
\begin{align}
    \Delta\theta_\vm = \Delta\varphi_\vm - \Delta\varphi_\bus,\
    s\Delta\theta_\vm = \Delta\omega_\vm - \Delta\omega_\bus,\label{eq:relative_frequency_ss}
\end{align}
to describe the relative angle difference and its derivative.

%
%

\subsection{Stand-Alone VSG Dynamics}

We obtain a small-signal transfer function for Figure~\ref{fig:VSG_diagram} as $$[\Delta\omega_\vm \; \Delta v_\vm]^\top = \bm{G}_\vm (s) \Delta\bm{s}_\ov.$$
We first need to eliminate $\Delta v_\bus$ from~\eqref{eq:power_flow_ss} and $\Delta\omega_\bus$ from \eqref{eq:swing_eq_ss} to relate the internal VSG dynamics only to the perturbation variables of power: $\Delta p_\ov$ and $\Delta q_\ov$. 

To this end, the following is obtained by solving each row of \eqref{eq:power_flow_ss} for $\Delta\theta_\vm$ and eliminating $\Delta\theta_\vm$:  \looseness=-1
%
%
{\medmuskip=.12mu\thickmuskip=.12mu\thinmuskip=.12mu
\begin{equation*}
    \frac{\Delta p_\ov - K_{\vm,12}\Delta v_\vm - K_{\vm,13}\Delta v_\bus}{K_{\vm,11}}
    = \frac{\Delta q_\ov - K_{\vm,22}\Delta v_\vm - K_{\vm,23}\Delta v_\bus}{K_{\vm,21}}
    .
\end{equation*}
}
Solving for $\Delta v_\bus$ yields
\begin{align}
    \Delta v_\bus &= [
        A_\mt{p,v} \; A_\mt{q,v} \; A_\mt{v,v}
    ]
[\Delta p_\ov \; \Delta q_\ov \; \Delta v_\vm]^\top
    ,
    \label{eq:vbus_expression}
    \\
        A_\mt{p,v} &= -\frac{K_{\vm,21}}{K_{\vm,11}K_{\vm,23}-K_{\vm,13}K_{\vm,21}} , 
        \nonumber 
        \\
        A_\mt{q,v} &= \frac{K_{\vm,11}}{K_{\vm,11}K_{\vm,23}-K_{\vm,13}K_{\vm,21}} , 
        \nonumber 
        \\
        A_\mt{v,v} &= \frac{K_{\vm,12}K_{\vm,21}-K_{\vm,11}K_{\vm,22}}{K_{\vm,11}K_{\vm,23}-K_{\vm,13}K_{\vm,21}} .
        \nonumber
\end{align}

To eliminate $\Delta\omega_\bus$ from \eqref{eq:swing_eq_ss}, bring in the following from~\eqref{eq:power_flow_ss}: $$s\Delta p_\ov = 
    [
        K_{\vm,11} \; K_{\vm,12} \; K_{\vm,13}]
    s [\Delta\theta_\vm \; \Delta v_\vm \; \Delta v_\bus]^\top.$$
Inserting \eqref{eq:relative_frequency_ss} and \eqref{eq:vbus_expression} into the equation above, we get
\begin{multline*}
    s\Delta p_\ov = K_{\vm,11}\left(\Delta\omega_\vm-\Delta\omega_\bus\right) + K_{\vm,12}s\Delta v_\vm 
    \\+ K_{\vm,13}\left(A_\mt{p,v}s\Delta p_\ov + A_\mt{q,v}s\Delta q_\ov + A_\mt{v,v}s\Delta v_\vm\right).
\end{multline*}
By reorganizing the terms above, we obtain
\begin{multline}
    \Delta\omega_\vm-\Delta\omega_\bus 
    = \frac{1-K_{\vm,13}A_\mt{p,v}}{K_{\vm,11}}s\Delta p_\ov 
    - \frac{K_{\vm,13}A_\mt{q,v}}{K_{\vm,11}}s\Delta q_\ov
    \\- \frac{K_{\vm,12}+K_{\vm,13}A_\mt{v,v}}{K_{\vm,11}}s\Delta v_\vm.\label{eq:wbus_subs}
\end{multline}
Towards our goal, we can now substitute \eqref{eq:wbus_subs} into \eqref{eq:swing_eq_ss}, yielding
{\medmuskip=.12mu\thickmuskip=.12mu\thinmuskip=.12mu
\begin{multline}
    M_\vm s\Delta\omega_\vm = 
    -\left(1 + D_\vm\frac{1-K_{\vm,13}A_\mt{p,v}}{K_{\vm,11}}s\right) \Delta p_\ov
    -\frac{K_\pv}{1+T_\pv s}\Delta\omega_\vm 
    \\
    +D_\vm\frac{K_{\vm,13}A_\mt{q,v}}{K_{\vm,11}}s \Delta q_\ov
    +D_\vm\frac{K_{\vm,12}+K_{\vm,13}A_\mt{v,v}}{K_{\vm,11}}s \Delta v_\vm.
    \nonumber
\end{multline}
}%
Finally, invoking \eqref{eq:voltage_dynamics_ss}, we get the following for $\Delta\omega_\vm$:
{
\medmuskip=.02mu\thickmuskip=.02mu\thinmuskip=.02mu
\begin{multline}
    \Delta\omega_\vm =
    -\frac{1 + \frac{D_\vm}{K_{\vm,11}}\left( 1-K_{\vm,13}A_\mt{p,v}\right) s}{M_\vm s + \frac{K_\pv}{1+T_\pv s}} \Delta p_\ov
    \\
    +\frac{\frac{D_\vm}{K_{\vm,11}}\left(K_{\vm,13}A_\mt{q,v}-\frac{K_\qv}{1+T_\qv s}\left( K_{\vm,12}+K_{\vm,13}A_\mt{v,v}\right)\right)}{M_\vm s + \frac{K_\pv}{1+T_\pv s}}s \Delta q_\ov
    .
    \nonumber
\end{multline}
}%

The voltage dynamics are governed by \eqref{eq:voltage_dynamics_ss}. Hence, the small-signal transfer function for stand-alone VSG operation is\looseness=-1
{
\medmuskip=.02mu\thickmuskip=.02mu\thinmuskip=.02mu
\begin{align}
    \bm{G}_\vm (s) &= \begin{bmatrix}
        G_{\vm,11}(s) & G_{\vm,12}(s) \\
        0 & -\frac{K_\qv}{1+T_\qv s}
    \end{bmatrix},
    \nonumber
    \\
    G_{\vm,11}(s) &= -\frac{1 + \frac{D_\vm}{K_{\vm,11}}\left( 1-K_{\vm,13}A_\mt{p,v}\right) s}{M_\vm s + \frac{K_\pv}{1+T_\pv s}}
    ,\nonumber\\
    G_{\vm,12}(s) &= \frac{\frac{D_\vm}{K_{\vm,11}}\left(K_{\vm,13}A_\mt{q,v}-\frac{K_\qv}{1+T_\qv s}\left( K_{\vm,12}+K_{\vm,13}A_\mt{v,v}\right)\right)}{M_\vm s + \frac{K_\pv}{1+T_\pv s}}s.\nonumber
\end{align}}%
Analogously, but with a different parametrization, $\bm{G}_\sg (s)$ can be obtained for the SG, satisfying $$[\Delta\omega_\sg \, \Delta v_\sg]^\top = \bm{G}_\sg (s) \Delta\bm{s}_\sg.$$ This is omitted in the interest of space.

%
%

\subsection{Interconnected System Dynamics}

Next, we derive the small-signal transfer function for the complete interconnected system as depicted in Figure~\ref{fig:system_diagram}.
These transfer functions are of the following form: $$\Delta\bm{y}_\vm = [\Delta\omega_\vm \; \Delta v_\vm]^\top
     = \bm{H}_{\load\to\vm}(s) \Delta\bm{s}_\load,$$ $$\Delta\bm{y}_\sg = [\Delta\omega_\sg \; \Delta v_\sg]^\top
     = \bm{H}_{\load\to\sg}(s) \Delta\bm{s}_\load.$$
Similar to the stand-alone operation, the derivation steps for VSG and SG are analogous. 
Given the focus of our sensitivity analysis, we present $\bm{H}_{\load\to\sg}(s)$.

Defining $\Delta\bm{y}_\bus = [\Delta\omega_\bus \, \Delta v_\bus]^\top$, our approach relies on an intermediate step by deriving 
$$\Delta\bm{s}_\ov = \bm{F}_\mt{b,v}(s) \Delta\bm{y}_\bus,$$
    $$\Delta\bm{s}_\os = \bm{F}_\mt{b,s}(s) \Delta\bm{y}_\bus,$$
to then isolate either of the VSG or SG dynamics via elimination of certain terms by substitution.

\subsubsection{Intermediate Step}
Multiplying both sides of \eqref{eq:power_flow_ss} by $s$, substituting~\eqref{eq:relative_frequency_ss}, and separating terms dependent on $\Delta\omega_\vm$ and $\Delta v_\vm$ from terms dependent on $\Delta\omega_\bus$ and $\Delta v_\bus$ results in 
{\medmuskip=.02mu\thickmuskip=.02mu\thinmuskip=.02mu
\begin{equation}
    s\Delta\bm{s}_\ov = \bm{K}_\vm \hspace{-.086cm}\begin{bmatrix}
        \Delta\omega_\vm - \Delta\omega_\bus \\
        s\Delta v_\vm \\
        s\Delta v_\bus
    \end{bmatrix}
    = \bm{B}_\mt{v,v}(s) \Delta\bm{y}_\vm + \bm{B}_\mt{b,v}(s) \Delta\bm{y}_\bus
    ,
    \label{eq:split_power_flow_VSG}
\end{equation}}
{\begin{align*}
    \bm{B}_\mt{v,v}(s) = \begin{bmatrix}
        K_{\vm,11} & K_{\vm,12}s \\
        K_{\vm,21} & K_{\vm,22}s
    \end{bmatrix},
    \bm{B}_\mt{b,v}(s) = \begin{bmatrix}
        -K_{\vm,11} & K_{\vm,13}s \\
        -K_{\vm,21} & K_{\vm,23}s
    \end{bmatrix}
    .
\end{align*}}For the SG, we have the following version of the above:
{\medmuskip=.02mu\thickmuskip=.02mu\thinmuskip=.02mu\begin{equation}
    s\Delta\bm{s}_\os = \bm{K}_\sg\hspace{.086cm} \begin{bmatrix}
        \Delta\omega_\sg - \Delta\omega_\bus \\
        s\Delta v_\sg \\
        s\Delta v_\bus
    \end{bmatrix}
    = \bm{B}_{\sg,\sg}(s)\Delta\bm{y}_\sg + \bm{B}_{\bus,\sg}(s)\Delta\bm{y}_\bus
    .
    \label{eq:split_power_flow_SG}
\end{equation}}%
$\bm{B}_\mt{s,s}(s)$ and $\bm{B}_\mt{b,s}(s)$ are omitted in the interest of space.

We can now obtain $\Delta\bm{s}_\ov$ 
in terms of $\Delta\bm{y}_\bus$. Inserting $\Delta\bm{y}_\vm = \bm{G}_\vm (s) \Delta\bm{s}_\ov$ into \eqref{eq:split_power_flow_VSG} and solving for $\Delta\bm{s}_\ov$ yields
\begin{equation}
    \Delta\bm{s}_\ov = \bm{F}_\mt{b,v}(s) \Delta\bm{y}_\bus
    ,
    \label{eq:bus_to_VSG_TF}
\end{equation}
where $\bm{F}_\mt{b,v}(s) = \left( s\bm{I} - \bm{B}_\mt{v,v}(s)\bm{G}_\vm (s) \right)^{-1} \bm{B}_\mt{b,v}(s).$
A similar expression can also be obtained for the SG as follows:
\begin{equation}
    \Delta\bm{s}_\os = \bm{F}_\mt{b,s}(s) \Delta\bm{y}_\bus,
    \label{eq:bus_to_SG_TF}
\end{equation}
where
$\bm{F}_\mt{b,s}(s) = \left( s\bm{I} - \bm{B}_\mt{s,s}(s)\bm{G}_\sg (s) \right)^{-1} \bm{B}_\mt{b,s}(s).$


\subsubsection{Complete Transfer Function for the SG}
We derive the small-signal transfer function $\bm{H}_{\load\to\sg}(s)$ from $\Delta\bm{s}_\load$ to~$\Delta\bm{y}_\sg$.
Inserting inverse of \eqref{eq:bus_to_SG_TF} into \eqref{eq:split_power_flow_SG} results in
    $$s\Delta\bm{s}_\os = \bm{B}_\mt{s,s}(s) \Delta\bm{y}_\sg + \bm{B}_\mt{b,s}(s) \bm{F}_\mt{b,s}^{-1}(s) \Delta\bm{s}_\os
    ,$$
and solving for $\Delta\bm{s}_\os$ gives
\begin{equation}
    \Delta\bm{s}_\os = \left( s\bm{I} - \bm{B}_\mt{b,s}(s) \bm{F}_{b,s}^{-1}(s)\right)^{-1}\bm{B}_\mt{s,s}(s) \Delta\bm{y}_\sg
    .
    \label{eq:eq:SG_TF_y_to_power}
\end{equation}

Notice that if we can relate $\Delta\bm{s}_\os$ to $\Delta\bm{s}_\load$, the derivation is complete.
To this end, insert \eqref{eq:bus_to_VSG_TF} and \eqref{eq:bus_to_SG_TF} into $\Delta\bm{s}_\load = \Delta\bm{s}_\ov + \Delta\bm{s}_\os,$ and then substitute inverse \eqref{eq:bus_to_SG_TF} to eliminate $\Delta\bm{y}_\bus$:
\begin{equation}
    \Delta\bm{s}_\load = \left( \bm{F}_\mt{b,v}(s) + \bm{F}_\mt{b,s}(s)\right)\bm{F}_\mt{b,s}^{-1}(s) \Delta\bm{s}_\os
    .
    \label{eq:power_balance_augmented}
\end{equation}

Combine \eqref{eq:eq:SG_TF_y_to_power} and \eqref{eq:power_balance_augmented} to obtain
$\Delta\bm{y}_\sg = \bm{H}_{\load\to\sg}(s) \Delta\bm{s}_\load$:
\begin{multline}
    \bm{H}_{\load\to\sg}(s) = \Bigl(
    \left( \bm{F}_\mt{b,v}(s) + \bm{F}_\mt{b,s}(s)\right)\bm{F}_\mt{b,s}^{-1}(s)
     \\ \times  
    \left( s\bm{I} - \bm{B}_\mt{b,s}(s)\bm{F}_\mt{b,s}^{-1}(s)\right)^{-1} \bm{B}_\mt{s,s}(s) 
    \Bigr)^{-1}.
    \label{eq:FULL_TF_to_SG}    
\end{multline}
\begin{remark}
    The same method also applies when deriving the transfer function $\bm{H}_{\load\to\vm}(s)$ for the VSG. The final result is
    \begin{multline*}
    \bm{H}_{\load\to\vm}(s) = \biggl(
    \left( \bm{F}_\mt{b,v}(s) + \bm{F}_\mt{b,s}(s)\right)\bm{F}_\mt{b,v}^{-1}(s)
    \\ 
    \times  
    \left( s\bm{I} - \bm{B}_\mt{b,v}(s)\bm{F}_\mt{b,v}^{-1}(s)\right)^{-1} \bm{B}_\mt{v,v}(s) 
    \biggr)^{-1}.
    \label{eq:FULL_TF_to_VSG}    
\end{multline*}
\end{remark}

\section{Sensitivity Analysis}
\label{ch:Sensitivity}

This section analyzes the sensitivity of SG dynamics in the interconnected system to variations of VSG parameters
All quantities are per-unit except time, frequency, and angles. Bases are:
nominal power of SG/VSG $S_\mt{n} = 1$~MVA, base voltage $V_\mt{n} = 6.6$~kV (phase-phase RMS), and nominal frequency $\omega_\mt{n} = 2\pi60$~rad/s, $\omega_\mt{r}=\omega_\mt{n}$. Parameters largely follow~\cite{liu2016vsg_small_signal}, except we adopt a large SG delay $T_{\mt{p},\sg}$ to reflect typically slow SG governors and prime movers~\cite{Lin2017,Liu2022}. This base case is listed in Table~\ref{tab:base_case_parameters}. The VSG’s default damper-winding constant is larger than the SG’s, and all remaining parameters match. As in~\cite{liu2016vsg_small_signal}, the SG and VSG include no stator resistance in the base case. Later, we vary the VSG's XR-ratio.

\begin{table}[t]
\vspace{0.1cm}
\centering
    \caption{Default parameters for the sensitivity analyses.}
    \vspace{0.1cm}
    \label{tab:base_case_parameters}
    \begin{tabular}{ l  c  c}
        \hline
          & VSG & SG \\
        \hline  
        Inertia constant $H_\cdot$ & $4.0$ s & $4.0$ s \\
        Damper winding constant $D_\cdot$ & $17$ pu & $3$ pu \\
        Frequency droop $K_\mt{p,\cdot}$ & $20$ pu & $20$ pu \\
        Governor lag $T_\mt{p,\cdot}$ & $1.0$ s & $1.0$ s \\
        QV droop $K_\mt{q,\cdot}$ & $0.1$ pu & $0.1$ pu \\
        QV droop lag $T_\mt{q,\cdot}$ & $0.1$ s & $0.1$ s \\
        Stator resistance $R_\cdot$ & $0.0$ pu & $0.0$ pu \\
        Stator reactance $X_\cdot$ & $0.2$ pu & $0.2$ pu \\
        \hline
    \end{tabular} 
    \vspace{0.1cm}
\end{table}

Power flow matrices $\bm{K}_\vm$ and $\bm{K}_\sg$ are dependent on the operating point $v_{\vm,0}$, $v_{\sg,0}$, $v_{\bus,0}$, $\theta_{\vm,0}$, and $\theta_{\sg,0}$, obtained by solving the power flow equations~\eqref{eq:power_flow_active_LS} and~\eqref{eq:power_flow_reactive_LS}. 
We set $p_{\ov,0} = p_{\os,0} = 0.5$~pu, $q_{\ov,0} = q_{\os,0} = 0.5$~pu, and $v_{\bus,0} = 1$~pu.
The resulting steady-state voltages and angles are in Table~\ref{tab:base_case_op}. 
Sensitivity results have also been checked at other set points with similar observations. All transfer functions are stable with large margins, so further details are omitted. 

\begin{table}[t]
\vspace{0.1cm}
\centering
    \caption{Operating point for the sensitivity analyses.}
    \vspace{0.1cm}
    \label{tab:base_case_op}
    \begin{tabular}{c  c c c}
         \hline  
         $v_{\vm,0}$ & $1.1045$~pu & $\theta_{\vm,0}$ & $0.0907$~rad \\
         $v_{\sg,0}$ & $1.1045$~pu & $\theta_{\sg,0}$ & $0.0907$~rad\\
         $v_{\bus,0}$ & $1.0000$~pu & &\\
        \hline
    \end{tabular} 
    \vspace{0.1cm}
\end{table}

\begin{figure}[t]
    \centering
    \includegraphics[width=\PlotScalingFactor\columnwidth]{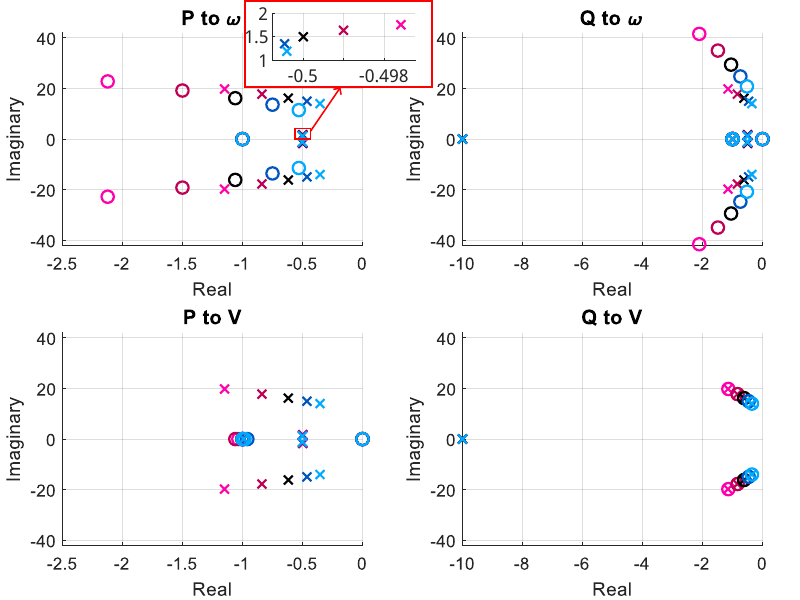}
    \caption{
        Pole-zero plots for inertia constants in the range $H_\vm \in \left[2, 8\right]$~s. 
        (\textbf{Black}: base case $H_\vm=4$~s, \textcolor{redFG}{\textbf{red}}: $H_\vm$ smaller than default, \textcolor{blueFG}{\textbf{blue}}: $H_\vm$ larger than default. Darker shades refer to parameter values closer to the base case. Crosses refer to poles, circles indicate zeros.)   
    }
    \label{fig:Sens_inertia_pz}
\end{figure}

\begin{figure}[t]
    \centering
    \includegraphics[width=\PlotScalingFactor\columnwidth]{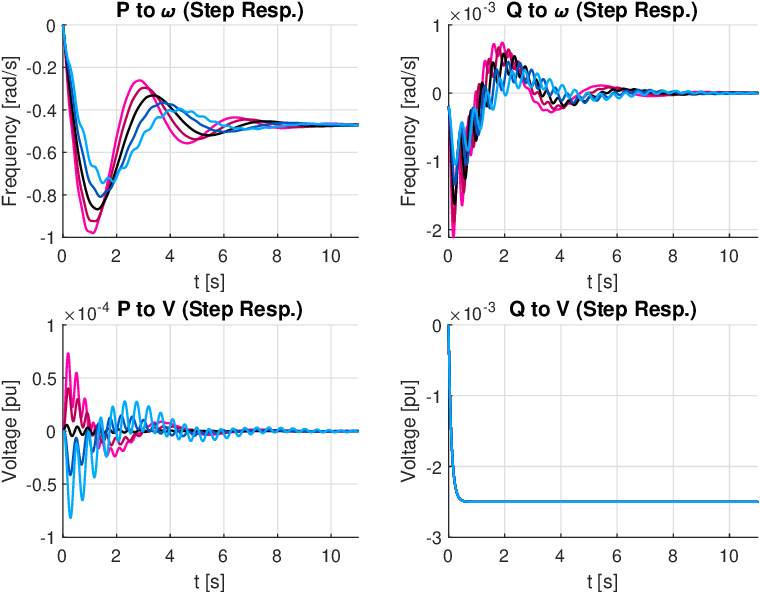}
    \caption{
        Step response plots for inertia constants in the range $H_\vm \in \left[2, 8\right]$~s. 
        (\textbf{Black}: base case $H_\vm=4$~s, \textcolor{redFG}{\textbf{red}}: $H_\vm$ smaller than default, \textcolor{blueFG}{\textbf{blue}}: $H_\vm$ larger than default. Darker shades refer to parameter values closer to the base case.)   
    }
    \label{fig:Sens_inertia_step}
\end{figure}

\subsection{Sensitivity With Respect to $H_\vm$}
\label{ch:Sens_Jv}

The inertia constant is varied within $\left[2,8\right]$~s. Fig.~\ref{fig:Sens_inertia_pz} shows poles and zeros for all four partial transfer functions of $\bm{H}_{\load\to\sg}(s)$ in \eqref{eq:FULL_TF_to_SG}. These four transfer functions are always referred to as \ptoomega{}, \qtoomega{}, \ptov{}, and \qtov{} in the following, i.e.,
$$\bm{H}_{\load\to\sg}(s)=\begin{bmatrix}
         P \to \omega & Q \to \omega\\
        P \to V & Q \to V
    \end{bmatrix}.$$

The base case in Table~\ref{tab:base_case_parameters} is plotted in black. Shades of red and blue denote smaller and larger inertia, respectively. Step responses are shown in Fig.~\ref{fig:Sens_inertia_step}. Step responses use $\Delta p_\load=\Delta q_\load=0.05$~pu. For visualizations, pole–zero pairs are canceled in plots using MATLAB’s minreal with tolerance $0.001$.

\subsubsection{Observations on the Base Case}
In the base case plotted in black,
\ptoomega{} and \qtoomega{}
show a slow resonant pole pair (here referred to as the \textit{primary} pole pair), resulting in a slow oscillatory behavior in the step response.
All four transfer functions also exhibit a fast resonant pole pair (here called the \textit{secondary} pole pair), which is partially canceled up to varying degree by accompanying zeros, except in \ptov{}.
Because of incomplete pole-zero cancellation, these faster oscillations are superimposed on the step response of \ptoomega{} and \qtoomega{}.
The cancellation is close to being perfect in \qtov{}.\looseness=-1


\subsubsection{Observations on Inertia}

As Figs.~\ref{fig:Sens_inertia_pz} and \ref{fig:Sens_inertia_step} show, increasing $H_\vm$ leads to a better damping ratio and a decline of the natural frequency of the primary pole pair for \ptoomega{} and \qtoomega{}, reconfirming related observations in~\cite{DArco2016,Lin2017,liu2016vsg_small_signal,leon2023virtual}.
For $H_\vm \neq H_\sg$, the primary pole pair is also present in \ptov{}.\looseness=-1

By contrast, the secondary poles shift to lower natural frequency and reduced damping for larger $H_\vm$, consistent with transient damping observations for inertialess droop controllers~\cite{he2021transient}. Zero–pole cancellation is best (in terms of natural frequency) when $H_\vm=H_\sg$. Accordingly, the step response of \ptoomega{} shows the smallest superimposed secondary oscillations at $H_\vm=H_\sg$. Thus, changes in pole–zero cancellation with $H_\vm$ can affect dynamics as much as pole locations.\looseness=-1

\qtoomega{} exhibits primary oscillations that decrease with larger $H_\vm$, similar to \ptoomega{}. Secondary oscillations are slower to decay as $H_\vm$ increases. \ptov{} contains large real zeros for $H_\vm\neq H_\sg$, which are omitted in Fig.~\ref{fig:Sens_inertia_pz} for clarity, and which grow in magnitude for increasing mismatch, yielding stronger primary and secondary oscillations. 

In \qtov{}, secondary resonant poles are effectively canceled and the real pole is insensitive to $H_\vm$. \qtoomega{} and \ptov{} are at least an order of magnitude smaller than \ptoomega{} and \qtov{} due to the base case not including stator resistances.\looseness=-1

%
%

\begin{figure}[t]
    \centering
    \includegraphics[width=\PlotScalingFactor\columnwidth]{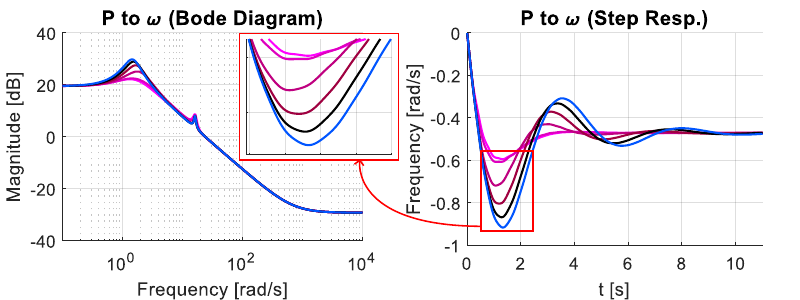}
    \caption{
        Bode magnitude plots and step responses of \ptoomega{} for $T_\pv\in\left[0.03,1.3\right]$~s.
        (\textbf{Black}: base case $T_\pv=1$~s, \textcolor{redFG}{\textbf{red}}: $T_\pv$ smaller than default, \textcolor{blueFG}{\textbf{blue}}: $T_\pv$ larger than default. Darker shades refer to parameter values closer to the base case.)
    }
    \label{fig:Sens_gov_lag}
\end{figure}

\subsection{Sensitivity With Respect to $T_\pv$}
\label{ch:Sens_Tpv}
Bode diagrams and step responses of \ptoomega{} for VSG governor-lag variations $T_\pv\in\left[0.03,1.3\right]$s (nominal $T_\pv=1$~s) are shown in Fig.~\ref{fig:Sens_gov_lag}.  Increasing $T_\pv$ strengthens primary oscillations, whereas small $T_\pv$ effectively damps them, consistent with~\cite[\S 3.C]{Liu2022},~\cite{liu2016vsg_small_signal}. Conversely, secondary oscillations decrease with larger $T_\pv$, revealing a trade-off. \qtov{} is largely unaffected by $T_\pv$, and is omitted.\looseness=-1

%
%

\subsection{Sensitivity With Respect to the XR-ratio}
\label{ch:Sens_XRv}
\begin{figure}[t]
    \centering
    \includegraphics[width=\PlotScalingFactor\columnwidth]{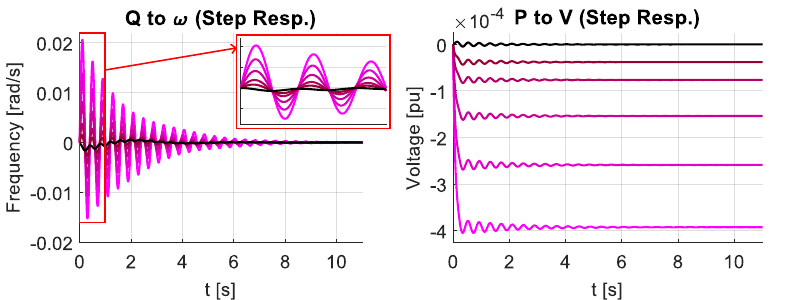}
    \caption{
        Step responses plots of \qtoomega{} and \ptov{} for various $X_\vm/R_\vm$-ratios.
        (\textbf{Black}: base case $X_\vm/R_\vm=\infty$, \textcolor{redFG}{\textbf{red}}: $X_\vm/R_\vm \in \left[2,20 \right]$. Darker shades refer to larger ratios.)
    }
    \label{fig:Sens_XR}
\end{figure}

Fig.~\ref{fig:Sens_XR} shows step responses of \qtoomega{} and \ptov{} for $X_\vm/R_\vm\in\{2,3,5,10,20,\infty\}$; $X_\vm$ is fixed at its default and $R_\vm$ is varied.
\ptoomega{} is unaffected over this range. Although not plotted, \qtov{} shows stronger secondary oscillations for smaller $X_\vm/R_\vm$. In general, \qtoomega{} and \ptov{} oscillations grow as $X_\vm/R_\vm$ decreases. For \qtoomega{}, secondary oscillations become dominant at small $X_\vm/R_\vm$, making primary oscillations negligible.  A similar observation has also been provided in~\cite[\S 3.C]{Liu2022} showing that decreasing the XR-ratio of VSG weakens the damping of low frequency oscillations. For \ptov{}, the XR-ratio also shifts the steady-state value and a smaller ratio yields slightly larger secondary oscillations.\looseness=-1

\subsection{Additional Results}
In this section, we list studies relegated to the appendix. Some of the main findings are later highlighted in Section~\ref{sec:conc}.
Appendix~\ref{app:A} provides Bode magnitude plots for sensitivity with respect to $H_\vm$ and an alternative base case with matched damper winding constants for the SG and VSG. It also includes sensitivities to $D_\vm$, $K_\qv$, $T_\qv$, and $X_\vm$.
Finally, Appendix~\ref{app:B} validates the analytical model of Section~\ref{ch:Dynamics} and the major observations via simulation case studies.

\section{Conclusion}\label{sec:conc}
We derived a small‑signal model to analyze how the VSG parameters influence the SG dynamics. Our main findings from the sensitivity analysis are:
\textit{(i)} For inertia, there is a trade-off between suppressing primary oscillations by choosing larger values and increasing secondary oscillations when deviating from $H_\vm=H_\sg$.
\textit{(ii)} A larger damper winding constant reduces secondary oscillations but is not effective in damping primary ones.
\textit{(iii)} Matched impedance for VSG and SG yields smaller secondary oscillations. 
\textit{(iv)} Smaller governor lag strongly reduces primary oscillations, even though this might induce slightly stronger secondary oscillations.
\textit{(v)} No negative effects from mismatch in QV-droop are observed. Stator impedance and QV-droop control together how the voltage reacts to a change in reactive power.
\textit{(vi)} The magnitude of oscillations grow with decreasing XR-ratio of VSG.


\bibliographystyle{IEEEtran}
\bibliography{bib_small_signal}

\begin{thebibliography}{10}
\providecommand{\url}[1]{#1}
\csname url@samestyle\endcsname
\providecommand{\newblock}{\relax}
\providecommand{\bibinfo}[2]{#2}
\providecommand{\BIBentrySTDinterwordspacing}{\spaceskip=0pt\relax}
\providecommand{\BIBentryALTinterwordstretchfactor}{4}
\providecommand{\BIBentryALTinterwordspacing}{\spaceskip=\fontdimen2\font plus
\BIBentryALTinterwordstretchfactor\fontdimen3\font minus
  \fontdimen4\font\relax}
\providecommand{\BIBforeignlanguage}[2]{{%
\expandafter\ifx\csname l@#1\endcsname\relax
\typeout{** WARNING: IEEEtran.bst: No hyphenation pattern has been}%
\typeout{** loaded for the language `#1'. Using the pattern for}%
\typeout{** the default language instead.}%
\else
\language=\csname l@#1\endcsname
\fi
#2}}
\providecommand{\BIBdecl}{\relax}
\BIBdecl

\bibitem{harnefors2015passivity}
L.~Harnefors, X.~Wang, A.~G. Yepes, and F.~Blaabjerg, ``Passivity-based
  stability assessment of grid-connected vscs—an overview,'' \emph{IEEE
  JESTPE}, vol.~4, no.~1, pp. 116--125, 2015.

\bibitem{Tayyebi2020}
A.~Tayyebi, D.~Groß, A.~Anta, F.~Kupzog, and F.~Dörfler, ``Frequency
  stability of synchronous machines and grid-forming power converters,''
  \emph{IEEE JESTPE}, vol.~8, no.~2, pp. 1004--1018, 2020.

\bibitem{sun2021stability}
P.~Sun, J.~Yao, Y.~Zhao, X.~Fang, and J.~Cao, ``Stability assessment and
  damping optimization control of multiple grid-connected virtual synchronous
  generators,'' \emph{IEEE Trans. on Energy Convers.}, vol.~36, no.~4, pp.
  3555--3567, 2021.

\bibitem{Liu2022}
H.~Liu, D.~Sun, P.~Song, X.~Cheng, F.~Zhao, and Y.~Tian, ``Influence of virtual
  synchronous generators on low frequency oscillations,'' \emph{CSEE J. of
  Power and Energy Sys.}, vol.~8, no.~4, pp. 1029--1038, 2022.

\bibitem{pogaku2007modeling}
N.~Pogaku, M.~Prodanovic, and T.~C. Green, ``Modeling, analysis and testing of
  autonomous operation of an inverter-based microgrid,'' \emph{IEEE Trans. on
  Power Elec.}, vol.~22, no.~2, pp. 613--625, 2007.

\bibitem{liu2016vsg_small_signal}
J.~Liu, Y.~Miura, and T.~Ise, ``Comparison of dynamic characteristics between
  virtual synchronous generator and droop control in inverter-based distributed
  generators,'' \emph{IEEE Trans. on Power Elec.}, vol.~31, no.~5, pp.
  3600--3611, 2016.

\bibitem{Lin2017}
Y.~Lin, B.~Johnson, V.~Gevorgian, V.~Purba, and S.~Dhople, ``Stability
  assessment of a system comprising a single machine and inverter with scalable
  ratings,'' in \emph{NAPS}, 2017, pp. 1--6.

\bibitem{lasseter2019grid}
R.~H. Lasseter, Z.~Chen, and D.~Pattabiraman, ``Grid-forming inverters: A
  critical asset for the power grid,'' \emph{IEEE JESTPE}, vol.~8, no.~2, pp.
  925--935, 2019.

\bibitem{Wu2016}
H.~Wu, X.~Ruan, D.~Yang, X.~Chen, W.~Zhao, Z.~Lv, and Q.-C. Zhong,
  ``Small-signal modeling and parameters design for virtual synchronous
  generators,'' \emph{IEEE Trans. on Ind. Elec.}, vol.~63, no.~7, pp.
  4292--4303, 2016.

\bibitem{DArco2016}
S.~D'Arco and J.~A. Suul, ``Small-signal analysis of an isolated power system
  controlled by a virtual synchronous machine,'' in \emph{IEEE PEMC}, 2016, pp.
  462--469.

\bibitem{li2016self}
D.~Li, Q.~Zhu, S.~Lin, and X.~Bian, ``A self-adaptive inertia and damping
  combination control of vsg to support frequency stability,'' \emph{IEEE
  Trans. on Energy Convers.}, vol.~32, no.~1, pp. 397--398, 2016.

\bibitem{leon2023virtual}
A.~E. Leon and J.~M. Mauricio, ``Virtual synchronous generator design to
  improve frequency support of converter-interfaced systems,'' \emph{IEEE
  Trans. on Energy Convers.}, 2024.

\bibitem{he2021transient}
X.~He, S.~Pan, and H.~Geng, ``Transient stability of hybrid power systems
  dominated by different types of grid-forming devices,'' \emph{IEEE Trans. on
  Energy Convers.}, vol.~37, no.~2, pp. 868--879, 2021.

\end{thebibliography}
\newpage
\appendices
\section{Additional Sensitivity Analyses}\label{app:A}

\subsection{Sensitivity With Respect to $H_\vm$}
\begin{figure}[t]
	\centering
	\includegraphics[width=\PlotScalingFactor\columnwidth]{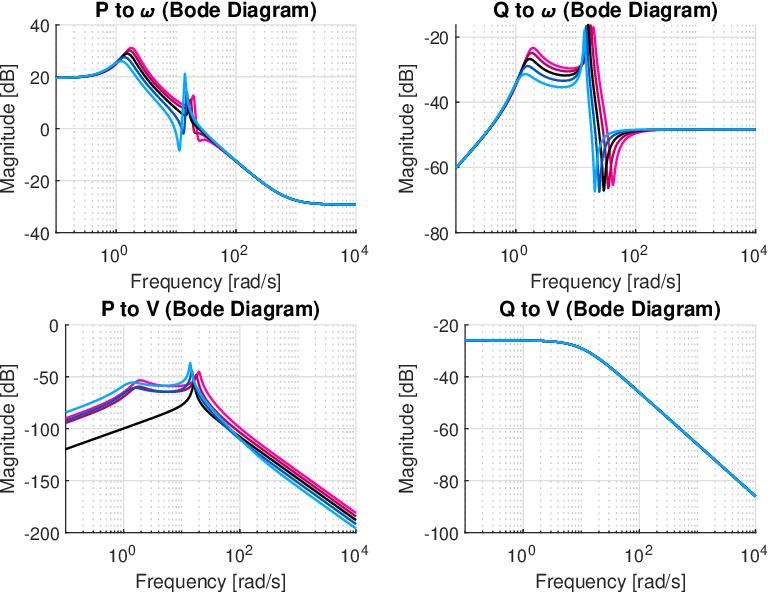}
	\caption{
		Bode magnitude plots for inertia constants in the range $H_\vm \in \left[2, 8\right]$~s. 
		(\textbf{Black}: base case $H_\vm=4$~s, \textcolor{redFG}{\textbf{red}}: $H_\vm$ smaller than default, \textcolor{blueFG}{\textbf{blue}}: $H_\vm$ larger than default. Darker shades refer to parameter values closer to the base case.)   
	}
	\label{fig:Sens_inertia_bode}
\end{figure}

Bode magnitude plots are provided in Fig.~\ref{fig:Sens_inertia_bode}.

\subsection{Base Case With Matching Damper Winding Constants}
\begin{figure}[t]
	\centering
	\includegraphics[width=\PlotScalingFactor\columnwidth]{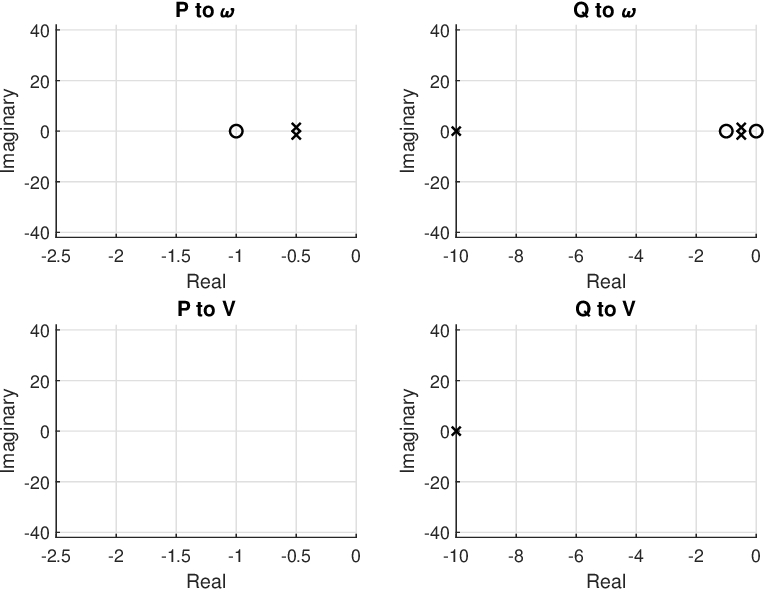}
	\caption{
		Pole-zero plots for a modified base case when the VSG parameters perfectly match the SG parameters.
		(\textbf{Black}: modified base case. Crosses refer to poles, circles indicate zeros.)   
	}
	\label{fig:Sens_all_match}
\end{figure}

When all the parameters including the damper winding constant are matched between the VSG and the SG, the secondary poles and zeros completely vanish from all partial transfer functions and \ptov{} fully reduces to $0$. The resulting pole-zero plots are provided in Fig.~\ref{fig:Sens_all_match}. The primary poles remain unaffected. Hence, the matched parameters case yields a less complex dynamics structure free from any secondary dynamics.

%
%

\subsection{Sensitivity With Respect to $D_\vm$}
\label{ch:Sens_Dv}

\begin{figure}[t]
	\centering
	\includegraphics[width=\PlotScalingFactor\columnwidth]{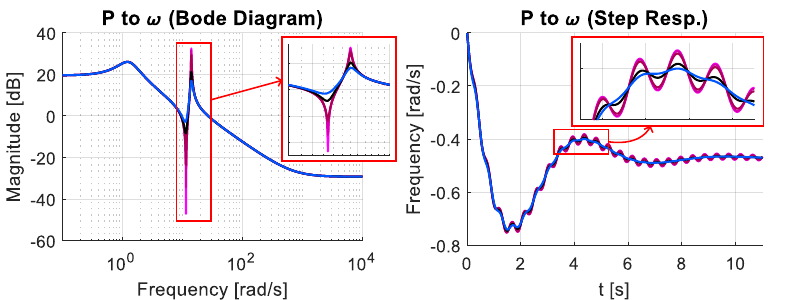}
	\caption{
		Bode magnitude plots and step responses of \ptoomega{} for $D_\vm\in\left[0.3,34\right]$~p.u with $H_\vm = 2 H_\vm^\mt{default}$ for all.
		(\textbf{Black}: base case $D_\vm=17$~pu, \textcolor{redFG}{\textbf{red}}: $D_\vm$ smaller than default, \textcolor{blueFG}{\textbf{blue}}: $D_\vm$ larger than default. Darker shades refer to parameter values closer to the base case.)
	}
	\label{fig:Sens_damping}
\end{figure}

Fig.~\ref{fig:Sens_inertia_step} shows that larger inertia yields stronger secondary oscillations. Hence, larger inertia $H_\vm = 2H_\vm^\mt{default}$ is chosen such that the effect of variations in $D_\vm$ is more pronounced.
\qtov{} is insensitive to changes in $D_\vm$. Thus,
Fig.~\ref{fig:Sens_damping} provides only the Bode plots and step responses of \ptoomega{} for damper winding constants in $\left[0.3,34\right]$~pu, where $D_\vm = 17$~pu by default.
$D_\vm$ has no effect on the primary resonant poles, and the primary oscillations remain intact.
The secondary oscillations decrease for larger damper winding constants. While the degree of pole-zero cancellation does not change with $D_\vm$, the pole damping increases with growing $D_\vm$.\looseness=-1

%
%

\subsection{Sensitivity With Respect to $K_\qv$}
\label{ch:Sens_Kqv}

\begin{figure}[t]
	\centering
	\includegraphics[width=\PlotScalingFactor\columnwidth]{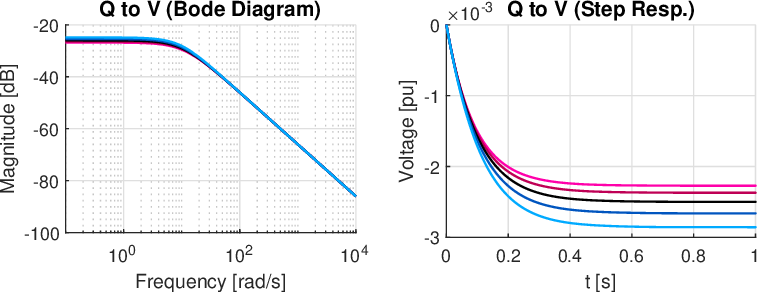}
	\caption{
		Bode magnitude plots and step responses of \qtov{} for $K_\qv\in\left[0.05,0.2\right]$~pu
		(\textbf{Black}: base case $K_\qv=0.1$~pu, \textcolor{redFG}{\textbf{red}}: $K_\qv$ smaller than default, \textcolor{blueFG}{\textbf{blue}}: $K_\qv$ larger than default. Darker shades refer to parameter values closer to the base case.)
	}
	\label{fig:Sens_qv_droop}
\end{figure}

We again use $H_\vm$ from Table~\ref{tab:base_case_parameters}. Varying $K_\qv$ within $\left[0.05,0.2\right]$~pu (default: $K_\qv = 0.1$~pu) has virtually no effect on the dynamics of \ptoomega{}, Thus, Fig.~\ref{fig:Sens_qv_droop} only shows \qtov{}. The dynamic behaviour of \qtov{} remains mostly unaffected by the changes in $K_\qv$, and only the steady-state value differs.
Thus, the choice of $K_\qv$ should depend on the desired droop level.

%
%

\subsection{Sensitivity With Respect to $T_\qv$}
\label{ch:Sens_Tqv}

\begin{figure}[t]
	\centering
	\includegraphics[width=\PlotScalingFactor\columnwidth]{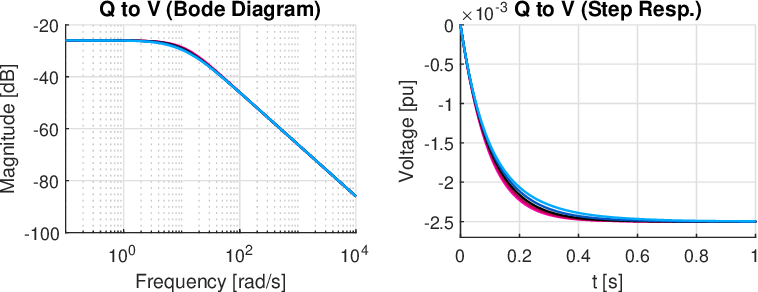}
	\caption{
		Bode magnitude plots and step responses of \qtov{} for $T_\qv\in\left[0.05,0.2\right]$~s.
		(\textbf{Black}: base case $T_\qv=0.1$~s, \textcolor{redFG}{\textbf{red}}: $T_\qv$ smaller than default, \textcolor{blueFG}{\textbf{blue}}: $T_\qv$ larger than default. Darker shades refer to parameter values closer to the base case.)
	}
	\label{fig:Sens_qv_droop_lag}
\end{figure}

Fig.~\ref{fig:Sens_qv_droop_lag} shows the Bode plots and step responses of \qtov{} for changing QV-droop time constants in the range $\left[0.05,0.2\right]$~s, where  $T_\qv = 0.1$~s by default. We observe that $T_\qv$ has no effect on \ptoomega{}. While not easily observed in the figure, \qtov{} contains a second real pole for $T_\qv \neq T_\mt{q,\sg}$, and for smaller lag $T_\qv$, a slightly faster convergence of the step response is observed. 

%
%

\subsection{Sensitivity With Respect to $X_\vm$}
\label{ch:Sens_Xv}

\begin{figure}[t]
	\centering
	\graphicspath{{Figures/Sensitivity/}}
	\includegraphics[width=\PlotScalingFactor\columnwidth]{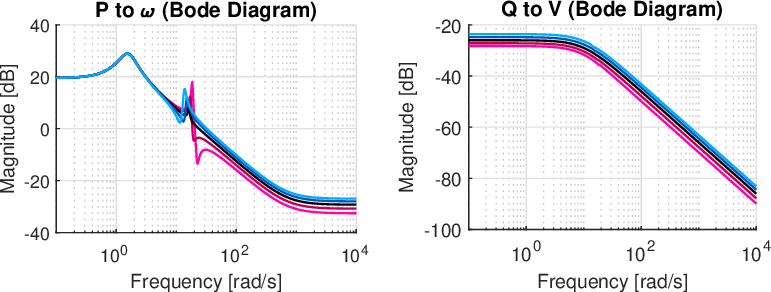}
	\caption{
		Bode magnitude plots of \ptoomega{} and \qtov{} for $X_\vm\in\left[0.1,0.4\right]$~pu
		(\textbf{Black}: base case $X_\vm=0.2$~pu, \textcolor{redFG}{\textbf{red}}: $X_\vm$ smaller than default, \textcolor{blueFG}{\textbf{blue}}: $X_\vm$ larger than default. Darker shades refer to parameter values closer to the base case.)
	}
	\label{fig:Sens_impedance}
\end{figure}

$X_\vm$ is varied within $\left[0.1, 0.4\right]$~pu, with the nominal value being $X_\vm = 0.2$~pu. The choice of $X_\vm$ also affects the operating point of the system, and we recompute it accordingly. The Bode plots of \ptoomega{} and \qtov{} are shown in Fig.~\ref{fig:Sens_impedance}.

While $X_\vm$ has no effect on the primary poles of \ptoomega{}, the location of its secondary poles and zeros changes, and the best pole-zero cancellation 
appears to be achieved for $X_\vm = X_\sg$.
Hence, matching impedances has significant importance.
In the mismatched case, an additional fast real pole appears in \qtov{}, and it becomes slower as $X_\vm$ increases. But even then, the main difference lies in this transfer function's steady-state value. As expected,
the change in voltage magnitude in response to the same step is larger for bigger $X_\vm$.

\subsection{Tuning Guidelines}
Here, we provide a set of additional numerical tuning guidelines for the VSG in the specific system under consideration based on the analytical models and previous observations:
\begin{itemize}
	\item[\textit{(i)}] Letting $H_\vm \in \left[4, 5\right]$~s ensures a good trade-off between the benefits of matching inertia and having slightly larger inertia.
	\item[\textit{(ii)}] Setting $D_\vm\geq17$~pu eliminates the secondary oscillations in \ptoomega{}.
	\item[\textit{(iii)}] Matching $X_\vm = X_\sg$, and keeping the XR-ratio of the VSG high are both indispensable.
	\item[\textit{(iv)}] Having a governor lag $T_\pv\in\left[0.06,0.12\right]$~s is desirable, as much lower values could create further oscillations. 
	\item[\textit{(v)}] Finally, QV-droop lag can be lower for the VSG, e.g, $T_\qv = 0.05$~s, to have slightly faster convergence.
\end{itemize}

\section{Simulation Case Studies}
\label{app:B}

This section verifies the analytical model derived in Section~\ref{ch:Dynamics} and some of the major observations with simulation case studies. 
We first explain the simulation setup and implementation to then provide step responses for active and reactive power steps. We vary $H_\vm$, $X_\vm$, XR-ratio, and the load impedance.
\looseness=-1

%
%

\begin{figure}[t]
    \centering
    \resizebox{.49\textwidth}{!}{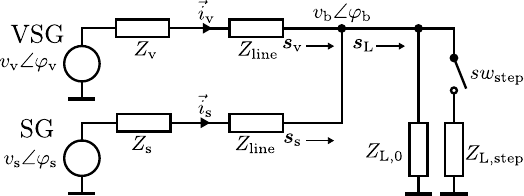}
    \caption{Diagram of the simulation model.}
    \label{fig:simulation_model}
\end{figure}

The system is implemented in MATLAB/Simulink. 
Fig.~\ref{fig:simulation_model} shows its diagram.
The grid model includes resistive-inductive line impedances.
Idealized controlled voltage sources with input signals $\vec{v}_\vm$ and $\vec{v}_\sg$ provide the internal voltages, assuming average converter models.
The load is modeled by the two impedances $Z_{\load,0}$ and $Z_{\load,\mt{step}}$, the values of which are determined to create the operating point and the power step. The step is performed by closing the switch $sw_\mt{step}$.

Unless later mentioned otherwise, we use the parameters in Table~\ref{tab:base_case_parameters} and the line impedances $Z_\mt{line} = 0.0054+0.0076j$~pu{, matching \cite{liu2016vsg_small_signal}}. The system is initialized to 
$\bm{s}_{\ov,\mt{r}} = \bm{s}_{\os,\mt{r}} = [0.5 \; 0.5]^\top$~pu
and the operating point described in Table~\ref{tab:base_case_op}. The impedance $Z_{\load,\mt{step}}$ is set to ensure a load step of 
$\Delta \bm{s}_\load = [0.02 \; 0.02]^\top$~pu.

Power flow matrices $\bm{K}_\vm$ and $\bm{K}_\sg$ depend on the voltage and angle operating points, and due to the line impedances, the real operating points slightly deviate from the precomputed solution of \eqref{eq:power_flow_active_LS} and \eqref{eq:power_flow_reactive_LS}. Hence, the analytical step responses are obtained using the actual operating points obtained from the simulations. 
Since $p_\load$ and $q_\load$ are perturbed simultaneously, these simulation studies do not identify \ptoomega{} and \qtoomega{} separately.
Hence, we focus on the signals $\Delta\omega_\sg$ and $\Delta v_\sg$.

%
%

\subsection{Verification of the Small-Signal Model}
\begin{figure}[t]
    \centering
    \includegraphics[width=\PlotScalingFactor\columnwidth]{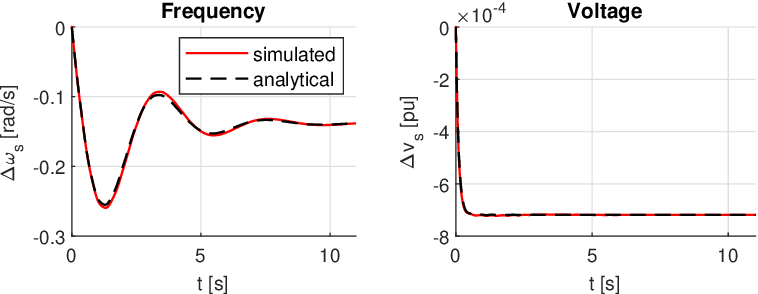}
    \caption{Comparison of simulated and analytical step responses (base case).}
    \label{fig:Sim_default}
\end{figure}

Fig.~\ref{fig:Sim_default} compares the step response computed from the small-signal model to the one obtained from the simulations. The parameters are as described above, and
$Z_{\load,0} = 0.5+0.5j$~pu and $Z_{\load,\mt{step}} = 25+25j$~pu accordingly. 
As Fig.~\ref{fig:Sim_default} shows, despite a minor mismatch during the transient, the simulation plots closely follow the predictions.

%
%

\subsection{Case Studies for Different Parameter Realisations}

\begin{figure}[t]
    \centering
    \includegraphics[width=\PlotScalingFactor\columnwidth]{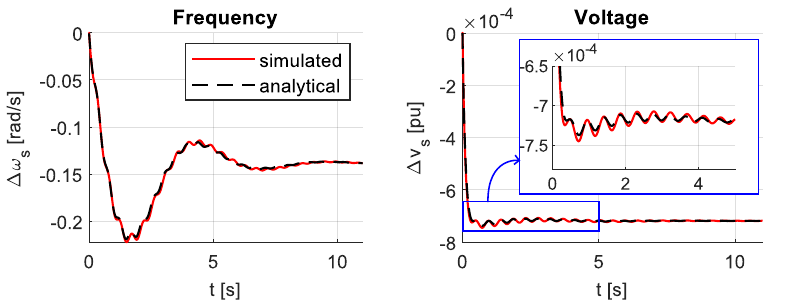}
    \caption{Comparison of step responses for $H_\vm = 2H_\vm^\mt{default}$.}
    \label{fig:sim_Jv}
\end{figure}

\begin{figure}[t]
    \centering
    \includegraphics[width=\PlotScalingFactor\columnwidth]{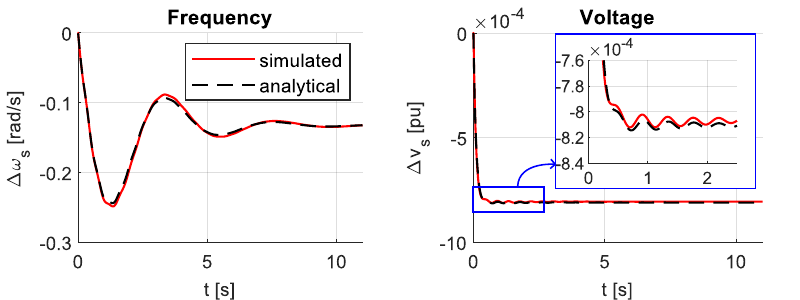}
    \caption{Comparison of step responses for $X_\vm = 1.4X_\vm^\mt{default}$.}
    \label{fig:sim_Xv}
\end{figure}

\begin{figure}[t]
    \centering
    \includegraphics[width=\PlotScalingFactor\columnwidth]{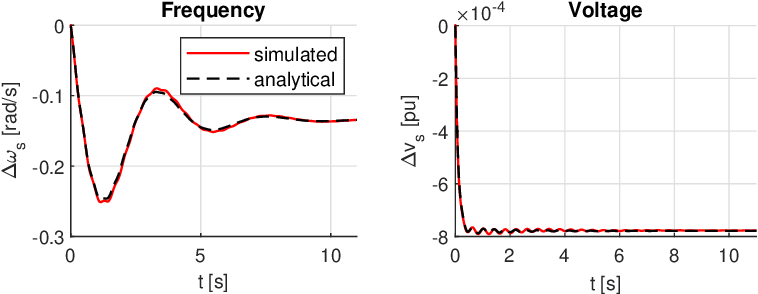}
    \caption{Comparison of step responses for $R_\vm \neq 0$ pu such that $X_\vm/R_\vm = 3$.}
    \label{fig:sim_XRv}
\end{figure}

\begin{figure}[t]
    \centering
    \includegraphics[width=\PlotScalingFactor\columnwidth]{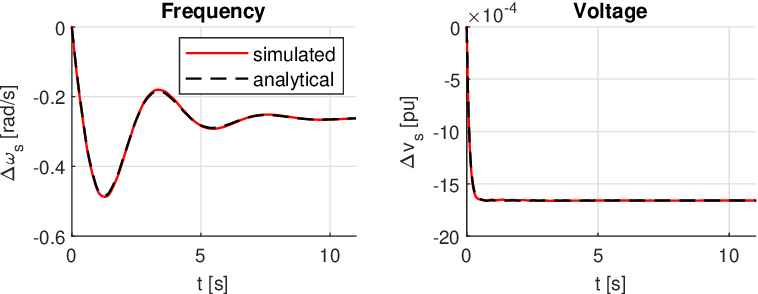}
    \caption{Comparison of step responses for changed power levels and loads: $\bm{s}_{\ov,r} = \bm{s}_{\os,r} = [0.5 \; 0.25]^\top$~pu and $\Delta \bm{s}_\load = [0.04 \; 0.04]^\top$~pu, yielding $Z_{\load,0} = 0.8 + 0.4j$~pu and $Z_{\load,\mt{step}} = 12.5+12.5j$~pu}
    \label{fig:sim_OP}
\end{figure}

We now vary the parameters.
First, Fig.~\ref{fig:sim_Jv} shows the case where $H_\vm = 2H_\vm^\mt{default}$. Second, the step responses for $X_\vm = 1.4X_\vm^\mt{default}$ are presented in Fig.~\ref{fig:sim_Xv}.
Third, Fig.~\ref{fig:sim_XRv} includes a nonzero stator resistance for the VSG: $X_\vm/R_\vm = 3$.
 Finally, we showcase the use of different load impedances in Fig.~\ref{fig:sim_OP}:
$\bm{s}_{\ov,r} = \bm{s}_{\os,r} = [0.5 \; 0.25]^\top$~pu
is chosen, such that $Z_{\load,0} = 0.8 + 0.4j$~pu. Furthermore, 
$\Delta \bm{s}_\load = [0.04 \; 0.04]^\top$~pu
yields $Z_{\load,\mt{step}} = 12.5+12.5j$~pu.
Setpoints, system initialisation, and power flow matrices are adjusted accordingly.


In all cases, the simulations support the analytical results.
Fig.~\ref{fig:sim_Jv} shows smaller primary oscillations and stronger secondary oscillations in the frequency step response than in Fig.~\ref{fig:Sim_default}, matching the findings from Section~\ref{ch:Sens_Jv}. Also, stronger secondary oscillations that can be attributed to \ptov{} can be observed in the combined voltage step response. 
Similarly, the slightly stronger oscillations in the frequency step response of Fig.~\ref{fig:sim_Xv} and the larger voltage drop in steady-state due $X_\vm > X_\sg$ support the analytical results from Section~\ref{ch:Sens_Xv}. As discussed in Section~\ref{ch:Sens_XRv}, due to a low XR-ratio, we can observe stronger secondary oscillations in both, the frequency and the voltage step response of Fig.~\ref{fig:sim_XRv} than in the base case.

In addition to that, Fig.~\ref{fig:sim_OP} illustrates that simulations and analytical results also match for different loads and power settings. We observe larger perturbations in frequency and voltage due to the bigger load step size.

\end{document}